\documentclass[amsmath, pra, twocolumn, showpacs, floatfix]{revtex4-1}
\usepackage{graphicx}
\usepackage{dsfont}

\begin{document} 

\title{Chiral excitation of a single atom by a single photon in a guided mode of a nanofiber}
 
\author{Fam Le Kien}
\affiliation{Okinawa Institute of Science and Technology Graduate University, Onna, Okinawa 904-0495, Japan}

\author{S\'{i}le Nic Chormaic}
\affiliation{Okinawa Institute of Science and Technology Graduate University, Onna, Okinawa 904-0495, Japan}

\author{Thomas Busch}
\affiliation{Okinawa Institute of Science and Technology Graduate University, Onna, Okinawa 904-0495, Japan}

\date{\today}

\begin{abstract}
We study the interaction between a single two-level atom and a single-photon probe pulse in a guided mode of a nanofiber. We examine the situation of chiral interaction, where the atom has a dipole rotating in the meridional plane of the nanofiber, and the probe pulse is quasilinearly polarized along the radial direction of the atom position in the fiber transverse plane. We show that the atomic excitation probability, the photon transmission flux, and the photon transmission probability depend on the propagation direction of the probe pulse along the fiber axis. 
In contrast, the reflection flux and the reflection probability do not depend on the propagation direction of the probe pulse. 
We find that the asymmetry parameter for the atomic excitation probability does not vary in time and does not depend on the probe pulse shape.
\end{abstract}

\maketitle

\section{Introduction}

The manipulation and control of coupling between light and matter at a single quantum level lie at the heart of quantum optics and quantum information processing and, therefore, have received a lot of attention in the past \cite{Cirac1997,Haroche1997,Duan2001,Dayan2014}. The interaction between a single two-level atom and a quantized single-photon light pulse has been studied extensively \cite{Domokos2002,Leuchs2007,Leuchs2009,Fan2010,Wang2011,Koenderink2011,Wang2012,GB}. It has been shown that the transient excitation probability of a single two-level atom interacting with a quantized single-photon pulse can achieve higher values than that in the steady-state regime. In particular, it has been predicted that the excitation probability of the atom can, in principle, approach unity if the photon waveform matches both spatially and temporally the time-reversed version of a spontaneously emitted photon \cite{Leuchs2007,Leuchs2009,Fan2010}. This condition means that the spatial profile of the incident photon should match the atomic dipole emission pattern and that the temporal shape of the incident photon should be a rising exponential \cite{Leuchs2007,Leuchs2009,Fan2010}. Schemes for efficient excitation involving free-space interaction \cite{Leuchs2007,Leuchs2009} as well as waveguides \cite{Fan2010,Wang2011,Koenderink2011,Wang2012,GB} have been studied. The analogy between a single atom and an optical resonator in the absorption of a light pulse has been investigated \cite{Leuchs2010,Leuchs2013}. Experiments on the use of rising exponential pulses for efficient atomic excitation, photon absorption, and loading of photons into a cavity at a single quantum level have been reported \cite{Du2012,Kurtsiefer2013,Martinis2014,Du2014,Kurtsiefer2016}.

It is difficult to achieve spatial mode matching between the incident photon wave packet and the atomic dipole emission profile when the atom is in free space. In contrast, the use of a waveguide provides strong spatial mode matching and hence simplifies practical implementations \cite{Fan2010,Wang2011,Koenderink2011,Wang2012,GB}. This strong mode matching is also the source of efficient channeling of spontaneous emission from atoms into fibers \cite{Jhe,Klimov,cesium decay}.

The efficient coupling between atoms and light can be seen clearly in nanofiber-based systems. Nanofibers are vacuum-clad, ultrathin optical fibers that allow tightly radially confined light to propagate over a long distance (the range of several millimeters is typical) and to interact efficiently with nearby atoms \cite{TongNat03,review2016,review2017,Nayak2018}. 
It has been shown that, for atoms near a nanofiber, spontaneous emission may become asymmetric with respect to opposite propagation directions \cite{Fam2014,Petersen2014,Mitsch14b,sponhigh}. This directional effect is a signature of spin-orbit coupling of light carrying transverse spin angular momentum \cite{Zeldovich,Bliokh review,Bliokh2014,Bliokh review2015,Bliokh2015,Banzer review2015,Lodahl2017}. The chirality of the field in a nanofiber-guided mode occurs as a consequence of the fact that the field has a nonzero longitudinal component, which oscillates in phase quadrature with respect to the radial transverse component. The chiral interaction of the guided field with a nearby atom appears when the atom has a dipole rotating in the meridional plane of the nanofiber. 

The purpose of this paper is to study the chiral interaction between a single two-level atom and a single-photon probe pulse in a guided mode of a nanofiber. 
We show that the atomic excitation probability, the photon transmission flux, and the photon transmission probability depend on the propagation direction of the probe pulse along the fiber axis. 

The paper is organized as follows. In Sec.~\ref{sec:model} we describe the model and the Hamiltonian of the system.
Section \ref{sec:equations} is devoted to the dynamical equations.
In Sec. \ref{sec:numerical}, we present the results of numerical calculations.
Our conclusions are given in Sec.~\ref{sec:summary}.

\section{Model and Hamiltonian}
\label{sec:model}

\begin{figure}[tbh]
	\begin{center}
		\includegraphics{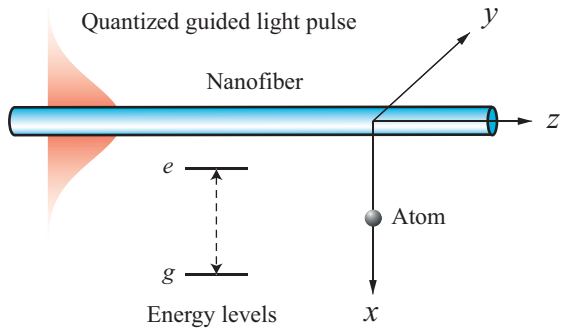}
	\end{center}
	\caption{Two-level atom interacting with a quantized light pulse in a guided mode of an optical nanofiber. 
	}
	\label{fig1}
\end{figure}

We consider a single two-level atom interacting with an injected quantized near-resonant light pulse in a guided mode of a vacuum-clad, ultrathin optical fiber (see Fig.~\ref{fig1}). The atom has an upper energy level $|e\rangle$ and a lower energy level $|g\rangle$, with energies $\hbar\omega_e$ and $\hbar\omega_g$, respectively, and is located at a fixed point outside the fiber. We assume that the central frequency $\omega_L$ of the probe pulse is close to the transition frequency $\omega_0=\omega_e-\omega_g$ of the atom,
and the spectral pulse width is small compared to the optical frequency. 
The fiber is a dielectric cylinder of radius $a$ and refractive index $n_1>1$ and is surrounded by an infinite background vacuum or air medium of refractive index $n_2=1$. We are interested in vacuum-clad silica-core ultrathin fibers with diameters in the range of hundreds of nanometers, which can support only the fundamental HE$_{11}$ mode and a few higher-order modes in the optical region. Such optical fibers are usually called nanofibers \cite{TongNat03,review2016,review2017,Nayak2018}.
In view of the very low losses of silica in the wavelength range of interest, we neglect material absorption. 

We use Cartesian coordinates $\{x,y,z\}$, where $z$ is the coordinate along the fiber axis, and also cylindrical coordinates $\{r,\varphi,z\}$, where $r$ and $\varphi$ are the polar coordinates in the fiber transverse plane $xy$.
We assume that 
the atom is located at a point $\mathbf{R}\equiv (r,\varphi,z)$ in the cylindrical coordinates. We use the notation $\mathbf{r}=(r,\varphi)$ for the position of the atom in the fiber transverse plane. 

The atom interacts with the full quantum electromagnetic field, which includes the injected quantum field in the input mode and the vacuum quantum field in other modes. In the presence of the fiber, the quantum field can be decomposed into the contributions from guided and radiation modes \cite{fiber books}. 
In the interaction picture, the Hamiltonian for the atom-field interaction in the dipole and rotating-wave approximations can be written as \cite{cesium decay,sponhigh} 
\begin{equation}\label{v1}
\begin{split}
H_{\mathrm{int}}&=-i\hbar\sum_{\alpha=\mu,\nu}(G_{\alpha}\sigma^{\dagger} a_{\alpha}e^{-i(\omega-\omega_0)t}-\mbox{H.c.}).
\end{split}
\end{equation}
Here, $\sigma=|g\rangle\langle e|$ and $\sigma^\dagger=|e\rangle\langle g|$ are the atomic transition operators, $a_{\alpha}$ and $a_{\alpha}^\dagger$ are the photon operators, 
and $G_\alpha$ is the coupling coefficient for the interaction between the atom and the quantum field in mode $\alpha$. To describe the atom, we use not only the transition operators $\sigma$ and $\sigma^\dagger$ but also the operators $\sigma_{ee}=|e\rangle\langle e|$ and $\sigma_{gg}=|g\rangle\langle g|$ for the populations
of the excited and ground states, respectively, and the operator $\sigma_z=\sigma_{ee}-\sigma_{gg}$ for the level population difference.

In Eq.~(\ref{v1}), the notations $\alpha=\mu,\nu$ and $\sum_{\alpha}=\sum_{\mu}+\sum_{\nu}$ stand for the mode index and the mode summation. The index $\mu=(\omega \mathcal{N} f p)$ labels guided modes. Here, $\omega$ is the mode frequency, $\mathcal{N}=\mathrm{HE}_{lm}$, EH$_{lm}$, TE$_{0m}$, or TM$_{0m}$ is the mode type, with $l=1,2,\dots$ and $m=1,2,\dots$ being the azimuthal and radial mode orders, $f=\pm1$ denotes the positive or negative propagation direction along the fiber axis $z$, and $p=\pm1$ for HE and EH modes and $0$ for TE and TM modes is the phase circulation direction index \cite{fiber books}. The longitudinal propagation constant $\beta$ of a guided mode is determined by the fiber eigenvalue equation. Meanwhile, the index $\nu=(\omega \beta l p)$ labels radiation modes. Here, $\beta$ is the longitudinal propagation constant, $l=0,\pm1,\pm2,\dots$ is the mode order, and $p=+,-$ is the mode polarization index. The longitudinal propagation constant $\beta$ of a radiation mode of frequency $\omega$ can vary continuously, from $-kn_2$ to $+kn_2$ (with $k=\omega/c$). The notations $\sum_{\mu}=\sum_{\mathcal{N} fp}\int_0^{\infty}d\omega$ and $\sum_{\nu}=\sum_{lp}\int_0^{\infty}d\omega\int_{-kn_2}^{kn_2}d\beta$ denote the generalized summations over guided and radiation modes, respectively.

The expressions for the coupling coefficients $G_{\alpha}$ with $\alpha=\mu,\nu$ are given as \cite{cesium decay,sponhigh} 
\begin{eqnarray}\label{v2}
G_{\mu}&=&\sqrt{\frac{\omega\beta'}{4\pi\epsilon_0\hbar}}\;
(\mathbf{d}\cdot\mathbf{e}^{(\mu)})e^{i(f\beta z+pl\varphi)},\nonumber\\
G_{\nu}&=&\sqrt{\frac{\omega}{4\pi\epsilon_0\hbar}}\;
(\mathbf{d}\cdot\mathbf{e}^{(\nu)})e^{i(\beta z+l\varphi)},
\end{eqnarray}
where $\mathbf{e}^{(\mu)}=\mathbf{e}^{(\mu)}(\mathbf{r})$ and $\mathbf{e}^{(\nu)}=\mathbf{e}^{(\nu)}(\mathbf{r})$ are the normalized mode functions given in \cite{fiber books, sponhigh}, 
$\beta'$ is the derivative of $\beta$
with respect to $\omega$, 
and $\mathbf{d}$ is the dipole matrix element of the atom. In general, the dipole matrix element $\mathbf{d}$ can be a complex vector. 

\section{Dynamical equations}
\label{sec:equations}

In this section, we derive the dynamical equations for interaction between the atom and a quantized probe light pulse in a guided mode of the nanofiber. In this derivation, we closely follow the techniques of Refs.~\cite{Domokos2002,Wang2011,Wang2012,GB} and extend them to include the specific characteristics of the nanofiber. 

\subsection{Heisenberg-Langevin equation for the atom interacting with a quantized guided light pulse}

In this subsection, we extend the Weisskopf-Wigner theory \cite{Scully} to
describe the observables of the internal state of the atom interacting with a quantized guided light pulse of the nanofiber. 
We call $\mathcal{O}$ an arbitrary atomic operator. The Heisenberg equation for this operator is
\begin{equation}\label{v3}
\begin{split}
\dot{\mathcal{O}}&=\sum_{\alpha}(G_{\alpha}
[\sigma^{\dagger},\mathcal{O}] a_{\alpha}
e^{-i(\omega-\omega_0)t}\\
&\quad+G_{\alpha}^{*}a_{\alpha}^{\dagger}[\mathcal{O},\sigma]
e^{i(\omega-\omega_0)t}.
\end{split}
\end{equation}
Meanwhile, the Heisenberg equation for the photon annihilation operator $a_{\alpha}$ is
$\dot{a}_{\alpha}=G_{\alpha}^*\sigma e^{i(\omega-\omega_0)t}$.
When we integrate this equation, we obtain
\begin{equation}\label{v4}
\begin{split}
a_{\alpha}(t)&=a_{\alpha}(t_0)+G_{\alpha}^*\int\limits _{t_0}^t dt'\,
\sigma(t')e^{i(\omega-\omega_0)t'},
\end{split}
\end{equation}
where $t_0$ is the initial time.

We assume that the evolution time $t-t_0$ and the characteristic atomic lifetime $\tau_0$ are 
large compared to the atomic transition period $2\pi/\omega_0$. When the continuum of the guided and radiation modes is regular and broadband around the atomic frequency $\omega_0$, the Markov approximation $\sigma(t')=\sigma(t)$ can be applied to describe the back
action of the second term in Eq.~(\ref{v4}) on the atom. 
Under the condition $t-t_0\gg 2\pi/\omega_0$, 
we calculate the integral with respect to $t'$ in the limit $t-t_0\to\infty$.
We set aside the imaginary part of the integral,
which describes the frequency shift. Such a frequency
shift is usually small. We can effectively account for it by
incorporating it into the atomic frequency. With the above
approximations and procedures, we find
$a_{\alpha}(t)=a_{\alpha}(t_0)+\pi G_{\alpha}^*\sigma(t)\delta(\omega-\omega_0)$.
We insert this expression into Eq.~(\ref{v3}). Then, we obtain the following Heisenberg-Langevin equation: 
\begin{equation}\label{v6}
\begin{split}
\dot{\mathcal{O}}&=\sum_{\alpha}(G_{\alpha}
[\sigma^{\dagger},\mathcal{O}] a_{\alpha}(t_0)
e^{-i(\omega-\omega_0)t}\\
&\quad+G_{\alpha}^{*}a_{\alpha}^{\dagger}(t_0)[\mathcal{O},\sigma]
e^{i(\omega-\omega_0)t})\\
&\quad
+\frac{1}{2}\sum\gamma(
[\sigma^{\dagger},\mathcal{O}] \sigma
+\sigma^{\dagger}[\mathcal{O},\sigma])
+\xi_{\mathcal{O}}.
\end{split}
\end{equation}
Here, the coefficient
$\gamma=2\pi\sum_{\alpha=\mu,\nu}|G_{\alpha}|^2\delta(\omega-\omega_0)$
is the total spontaneous emission rate of the atom and $\xi_{\mathcal{O}}$ is the noise operator. Note that the total spontaneous emission rate $\gamma$ can be decomposed as $\gamma=\gamma_g+\gamma_r$,
where $\gamma_g=2\pi\sum_{\mu}|G_{\mu}|^2\delta(\omega-\omega_0)$ and
$\gamma_r=2\pi\sum_{\nu}|G_{\nu}|^2\delta(\omega-\omega_0)$ are the rates of spontaneous emission into guided and radiation modes, respectively. 

We assume that the initial field is a quantum pulse light field propagating in a superposition of guided modes $(\omega\mathcal{N}_Lf_Lp_L)$ with the frequency $\omega$ varying in a small interval around a central frequency $\omega_L$.
We introduce the label $\mu_L=(\mathcal{N}_Lf_Lp_L)$ for this integral mode.
When the bandwidth of the pulse is narrow and the field central frequency $\omega_L$ is close to the atomic transition frequency $\omega_0$, we can use the approximation 
$\sum_{\alpha}G_{\alpha}a_{\alpha}(t_0)e^{-i(\omega-\omega_0)t}\cong G_{L}\int_0^{\infty} a_{\omega}(t_0)e^{-i(\omega-\omega_0)(t-f_Lz/v_{g_L})}d\omega$. Here, $G_{L}=G_{\omega_0\mathcal{N}_Lf_Lp_L}$, $a_{\omega}=a_{\omega\mathcal{N}_Lf_Lp_L}$, and $v_{g_L}=1/\beta'_L(\omega_0)$ are the coupling coefficient, the photon operator, and the group velocity of the input guided mode, respectively. 
Then, we can rewrite Eq.~\eqref{v6} as
\begin{equation}\label{v8}
\begin{split}
\dot{\mathcal{O}}&=\sqrt{2\pi}(G_{L}
[\sigma^{\dagger},\mathcal{O}] a_{t_d}+G_{L}^{*}a_{t_d}^{\dagger} [\mathcal{O},\sigma])\\
&\quad+\frac{1}{2}\gamma(
[\sigma^{\dagger},\mathcal{O}] \sigma
+\sigma^{\dagger}[\mathcal{O},\sigma])
+\xi_{\mathcal{O}},
\end{split}
\end{equation}
where $t_d=t-f_Lz/v_{g_L}$ and
\begin{equation}\label{v9}
a_t=\frac{1}{\sqrt{2\pi}}\int_0^{\infty} a_{\omega}(t_0) e^{-i(\omega-\omega_0)t}d\omega.
\end{equation}
We note that Eq.~(\ref{v8}) is in agreement with Eqs.~(13) and (14) of Ref.~\cite{Wang2012}. 

In deriving Eq.~(\ref{v8}), we have used the mode function for the quasicircularly polarized mode $\mu_L=(\mathcal{N}_Lf_Lp_L)$ to describe the input field. However, this equation can also be used for the quasilinearly polarized mode $\mu_L=(\mathcal{N}_Lf_L\varphi_{\mathrm{pol}})$, where the angle $\varphi_{\mathrm{pol}}$ characterizes the orientation of the principal polarization axis in the fiber transverse plane $xy$. 
For this mode, the coupling coefficient is given as $G_L=(e^{-i\varphi_{\mathrm{pol}}}G_{\omega_0\mathcal{N}_Lf_L,p_L=+}+e^{i\varphi_{\mathrm{pol}}}G_{\omega_0\mathcal{N}_Lf_L,p_L=-})/\sqrt2$. Note that the rate of spontaneous emission from the atom into the input guided mode $\mu_L$ is $\gamma_L=2\pi |G_L|^2$. This rate characterizes the strength of the coupling between the atom and the input field. The coupling efficiency is characterized by the parameter $\eta_L=\gamma_L/\gamma$.

\subsection{Quantized light pulses}

Quantized light pulses are described by the continuous-mode quantization formalism \cite{Loudon}. We briefly summarize below the key points of this description \cite{Loudon,Wang2011,Wang2012}.

A quantized light pulse can be considered as a photon wave packet. 
The photon wave-packet creation operator is defined as
\cite{Loudon}
\begin{equation}\label{v10}
A^\dagger=\int_{-\infty}^\infty F_t a^\dagger_t dt=\int_{0}^\infty F_\omega a^\dagger_{\omega} d\omega,
\end{equation}
where $a^\dagger_t$ and $a^\dagger_\omega=a^\dagger_\omega(t_0)$ are the photon creation operators in the time and frequency domains, respectively, 
and $F_t$ and $F_\omega$ are the temporal shape and spectral distribution of the wave packet.
They are related by the Fourier transformation
\begin{equation}\label{v11}
\begin{split}
a_t&=\frac{1}{\sqrt{2\pi}}\int_{0}^\infty e^{-i(\omega-\omega_0) t} a_{\omega} d\omega,\\
F_t&=\frac{1}{\sqrt{2\pi}}\int_{0}^\infty e^{-i(\omega-\omega_0) t} F_\omega d\omega.
\end{split}
\end{equation}
The amplitudes $F_t$ and $F_\omega$ are normalized as
$\int_{-\infty}^\infty |F_t|^2 dt=\int_{0}^\infty |F_\omega|^2 d\omega=1$.

The Fock state of the wave packet with the photon number $n=0,1,2,\dots$ is defined as \cite{Loudon}
\begin{equation}\label{v13}
|n\rangle=\frac{1}{\sqrt{n!}} (A^{\dagger})^n |0\rangle.
\end{equation}
The Fock state $|n\rangle$ has the properties $a_t|n\rangle=\sqrt{n} F_t|n-1\rangle$, $a_\omega|n\rangle=\sqrt{n} F_\omega|n-1\rangle$, 
$A|n\rangle=\sqrt{n}|n-1\rangle$, and $A^\dagger|n\rangle=\sqrt{n+1}|n+1\rangle$.

The coherent state of the wave packet with the complex amplitude $\alpha$ is defined as \cite{Loudon}
\begin{equation}\label{v16}
|\alpha\rangle=e^{-|\alpha|^2/2}\sum_n \frac{\alpha^n}{\sqrt{n!}}|n\rangle.
\end{equation}
It has the properties $a_t|\alpha\rangle=\alpha F_t|\alpha\rangle$, $a_\omega|\alpha\rangle=\alpha F_\omega|\alpha\rangle$, and $A|\alpha\rangle=\alpha|\alpha\rangle$. 

In the continuous-mode quantization formalism, the photon number operator is defined as $\hat{n}=\int_{-\infty}^\infty a^\dagger_t a_t dt=\int_{0}^\infty a^\dagger_{\omega} a_{\omega} d\omega$ \cite{Loudon}. We have $\hat{n}|n\rangle=n|n\rangle$ and
$\langle\alpha|\hat{n}|\alpha\rangle=|\alpha|^2$.

\subsection{Interaction of the atom with a Fock- or coherent-state pulse}

In this subsection, we derive the dynamical equations for the atom interacting with a Fock- or coherent-state light pulse.

First, we consider the interaction of the atom with a Fock-state pulse of $N$ photons. 
We assume that the atom is initially in the ground state $|g\rangle$.
We introduce the notation $|g,n,0\rangle$ for the state where the atom is in the ground state with
$n$ photons in the pulse field and no photons in the other modes. 
We also introduce the notation $\langle\mathcal{O}\rangle_{nn'}=\langle g,n,0|\mathcal{O}|g,n',0\rangle$. Without loss of generality, we assume that the axial coordinate of the atom is $z=0$. In this case, we have $t_d=t$.
Then, Eq.~(\ref{v8}) yields \cite{Wang2012}
\begin{eqnarray}\label{v19}
\langle\dot{\sigma}_z\rangle_{nn'}&=&-\gamma(\langle\sigma_z\rangle_{nn'}+\delta_{nn'})
-2\sqrt{2\pi n'}G_{L} F_t \langle\sigma^\dagger\rangle_{n,n'-1}\nonumber\\
&&\mbox{} -2\sqrt{2\pi n}G_{L}^{*}F_t^*\langle\sigma\rangle_{n-1,n'},\nonumber\\
\langle\dot{\sigma}\rangle_{nn'}&=&-\frac{\gamma}{2} \langle\sigma\rangle_{nn'}
+\sqrt{2\pi n'} G_{L}F_t\langle\sigma_z\rangle_{n,n'-1},
\end{eqnarray}
where $n$ and $n'$ run from 0 to $N$.
The initial conditions are $\langle\sigma_z(t_0)\rangle_{nn'}=-\delta_{nn'}$ and $\langle\sigma(t_0)\rangle_{nn'}=0$.
It follows from these initial conditions and Eqs.~(\ref{v19}) that the only nonzero matrix elements of the atomic operators
are $\langle\sigma_z\rangle_{nn}$, $\langle\sigma^\dagger\rangle_{n,n-1}$, and $\langle\sigma\rangle_{n-1,n}$. The time dependencies of these matrix elements are governed by the coupled equations \cite{Wang2012}
\begin{eqnarray}\label{v20}
\langle\dot{\sigma}_z\rangle_{nn}&=&-\gamma(\langle\sigma_z\rangle_{nn}+1)
-2\sqrt{2\pi n}G_{L} F_t \langle\sigma^\dagger\rangle_{n,n-1}\nonumber\\
&&\mbox{} -2\sqrt{2\pi n}G_{L}^{*}F_t^*\langle\sigma\rangle_{n-1,n},\nonumber\\
\langle\dot{\sigma}\rangle_{n-1,n}&=&-\frac{\gamma}{2} \langle\sigma\rangle_{n-1,n}
+\sqrt{2\pi n} G_{L}F_t\langle\sigma_z\rangle_{n-1,n-1},\qquad
\end{eqnarray}
where $n$ runs from 1 to $N$.
The corresponding initial conditions are $\langle\sigma_z(t_0)\rangle_{nn}=-1$ and $\langle\sigma(t_0)\rangle_{n,n+1}=0$.
Note that $\langle\sigma_z(t)\rangle_{00}=-1$ for any $t\ge t_0$.

We now consider the interaction of the atom with a pulse in a coherent state $\alpha$. 
We introduce the notations
$\langle\sigma_z\rangle=\langle g,\alpha,0|\sigma_z|g,\alpha,0\rangle$ and
$\langle\sigma\rangle=\langle g,\alpha,0|\sigma|g,\alpha,0\rangle$.
Then, Eq.~(\ref{v8}) yields \cite{Wang2011,Wang2012}
\begin{eqnarray}\label{v22}
\langle\dot{\sigma}_z\rangle&=&-\gamma (\langle\sigma_z\rangle+1)
-2\sqrt{2\pi}\alpha G_{L} F_t \langle\sigma^\dagger\rangle
\nonumber\\&&\mbox{} 
-2\sqrt{2\pi}\alpha^* G_{L}^{*}F_t^*\langle\sigma\rangle,\nonumber\\
\langle\dot{\sigma}\rangle&=&-\frac{\gamma}{2} \langle\sigma\rangle
+\sqrt{2\pi}\alpha G_{L}F_t\langle\sigma_z\rangle.
\end{eqnarray}
Note that Eqs.~(\ref{v22}) are the same as the equations for a two-level atom interacting with a classical driving field.

\subsection{Interaction of the atom with a single-photon Fock-state pulse}

In this subsection, we consider the case of a single-photon Fock-state pulse, that is, the case where the pulse is initially in the Fock state $|N\rangle$ with the photon number $N=1$. In this case, Eqs.~(\ref{v20}) reduce to
\cite{Wang2011,Wang2012}
\begin{eqnarray}\label{v23}
\dot{P}&=&-\gamma P
-\sqrt{2\pi}G_{L} F_t Q^*
-\sqrt{2\pi}G_{L}^{*}F_t^*Q,\nonumber\\
\dot{Q}&=&-\frac{\gamma}{2} Q -\sqrt{2\pi} G_{L}F_t,
\end{eqnarray}
where $P=(1+\langle g,1,0|\sigma_z|g,1,0\rangle)/2$ and
$Q=\langle g,0,0|\sigma|g,1,0\rangle$,
with the initial conditions $P(t_0)=0$ and $Q(t_0)=0$.
The quantities $P$ and $Q$ are the excitation probability and the induced dipole amplitude, respectively, of the atom. The solution of Eqs.~(\ref{v23}) for $t\geq t_0$ reads \cite{GB}
\begin{equation}\label{v25a}
P=2\pi|G_{L}|^2\left|\int_{t_0}^t e^{-\gamma(t-t')/2} F_{t'}dt' \right|^2
\end{equation}
and
\begin{equation}\label{v25b}
Q=-\sqrt{2\pi} G_{L}\int_{t_0}^t e^{-\gamma (t-t')/2} F_{t'} dt'.
\end{equation}
It is clear that $P=|Q|^2$. Note that, in the case of a coherent-state pulse with mean photon number $\bar{N}=|\alpha|^2=1$, Eqs.~(\ref{v22}) do not reduce to Eqs.~(\ref{v23}).
The two sets of equations agree with each other only in the case of $\langle\sigma_z\rangle\simeq-1$, that is, the case of weak atomic excitation. 

We note that the temporal shape of the single-photon probe pulse can be arbitrary and is described by the profile function $F_t$.
It has been shown in Refs.~\cite{Wang2011,Wang2012,GB} that the excitation of the atom depends on the temporal profile of the probe pulse. According to these references, the maximal value of the excitation probability $P$ is $P_{\mathrm{max}}=\eta_L=\gamma_L/\gamma=2\pi |G_L|^2/\gamma$. This value can be achieved at $t=0$ for a rising exponential resonant pulse, $F_t=T^{-1/2}e^{t/2T}$ for $t\leq0$ and 0 for $t>0$, with the time constant $T=1/\gamma$. It is worth mentioning here that the techniques for generation of single-photon pulses of various shapes have been demonstrated \cite{Du2012,Kurtsiefer2013,Martinis2014,Du2014,Kurtsiefer2016}. Below, we extend the treatment of Ref.~\cite{GB} and present the explicit analytical expressions for $P$ and $Q$ in the particular cases where the shape of the single-photon probe pulse is Gaussian, exponentially rising, or exponentially decaying, with a possible detuning $\Delta$.

\subsubsection{Gaussian pulse}

First, we consider the case of a Gaussian single-photon Fock-state pulse, where the pulse form function is $F_t=(2\pi T^2)^{-1/4}e^{-t^2/4T^2-i\Delta t}$. Here, $T$ is the characteristic pulse duration and $\Delta=\omega_L-\omega_0$ is the detuning of the field central frequency $\omega_L$ from the atomic transition frequency $\omega_0$. In this case, we find \cite{GB}
\begin{eqnarray}\label{v26}
P&=&
(\pi T^2/2)^{1/2}
2\pi|G_{L}|^2e^{-\gamma t+(\gamma^2-4\Delta^2)T^2/2}
\nonumber\\
&&\mbox{}\times
\Big|1+\mathrm{erf}\Big(\frac{t}{2T}-\frac{\gamma T}{2}+i\Delta T\Big)\Big|^2,\nonumber\\
Q&=& 
-(\pi T^2/2)^{1/4}\sqrt{2\pi} G_{L}e^{-\gamma t/2+(\gamma-2i\Delta)^2T^2/4}\nonumber\\
&&\mbox{}\times
\Big[1+\mathrm{erf}\Big(\frac{t}{2T}-\frac{\gamma T}{2}+i\Delta T\Big)\Big].
\end{eqnarray}

\subsubsection{Rising exponential pulse}

Next, we consider the case of a rising exponential single-photon Fock-state pulse, where the pulse form function is $F_t=T^{-1/2}e^{t/2T-i\Delta t}$ for $t\leq0$ and 0 for $t>0$. In this case, we find \cite{GB}
\begin{eqnarray}\label{v27}
P&=&
\frac{8\pi T}{(1+\gamma T)^2+4\Delta^2 T^2} |G_{L}|^2 e^{t/T},\nonumber\\
Q&=& 
-\frac{2\sqrt{T}}{1+\gamma T-2i\Delta T}\sqrt{2\pi} G_{L}e^{t/2T-i\Delta t}\qquad
\end{eqnarray}
for $t\leq0$, and 
\begin{eqnarray}\label{v28}
P&=&
\frac{8\pi T}{(1+\gamma T)^2+4\Delta^2 T^2} |G_{L}|^2e^{-\gamma t},\nonumber\\
Q&=& 
-\frac{2\sqrt{T}}{1+\gamma T-2i\Delta T}\sqrt{2\pi} G_{L}e^{-\gamma t/2}\qquad
\end{eqnarray}
for $t>0$.
It is clear that the maximal value of the excitation probability is $P_{\mathrm{max}}=2\pi |G_L|^2/\gamma=\gamma_L/\gamma$ and can be achieved at $t=0$ for a rising exponential resonant pulse with $T=1/\gamma$ and $\Delta=0$.

\subsubsection{Decaying exponential pulse}

Finally, we consider the case of a decaying exponential single-photon Fock-state pulse, where the pulse form function is
$F_t=T^{-1/2}e^{-t/2T-i\Delta t}$ for $t\geq0$ and 0 for $t<0$. In this case, we find \cite{GB}
\begin{eqnarray}\label{v29}
P&=&
\frac{8\pi T}{(1-\gamma T)^2+4\Delta^2 T^2} |G_{L}|^2 (e^{-t/T}+e^{-\gamma t}\nonumber\\
&&\mbox{}
-2e^{-t/2T-\gamma t/2}\cos\Delta t),\nonumber\\
Q&=&\frac{2\sqrt{T}}{1-\gamma T+2i\Delta T}\sqrt{2\pi} G_{L}(e^{-t/2T-i\Delta t}-e^{-\gamma t/2})\qquad
\end{eqnarray}
for $t\geq0$, and 
$P=Q=0$
for $t<0$.

We note that, in the case where $\Delta=0$, Eqs.~(\ref{v26})--(\ref{v29}) reduce to the results of Ref.~\cite{GB}.

\subsection{Photon transmission and reflection fluxes}

In this subsection, we derive the expressions for the fluxes of transmitted and reflected photons. 

In the framework of the continuous-mode quantization formalism,
the flux of photons in the guided modes propagating in the direction $f$ through the fiber cross-sectional plane at a position $z$ is given by \cite{Loudon}
\begin{equation}\label{v30}
I_f(z,t)=\sum_{\mathcal{N}p}\langle A_{\mathcal{N}fp}^\dagger(z,t) A_{\mathcal{N}fp}(z,t)\rangle,
\end{equation}
where
\begin{equation}\label{v31}
A_{\mathcal{N}fp}(z,t)=\frac{1}{\sqrt{2\pi}}
\int_0^{\infty}d\omega\, a_{\omega\mathcal{N}fp}(t) e^{-i(\omega t-f\beta z)}
\end{equation}
is the Fourier-transformed photon operator.

Let the atom be located at a point $\mathbf{R}_a=(r_a,\varphi_a,z_a)$. 
We insert Eq.~(\ref{v4}) into Eq.~(\ref{v31}).
Under the condition of narrow bandwidth,
we use the approximations $G_{\omega\mathcal{N} fp}(\mathbf{R}_a)=G_{\omega_0\mathcal{N} fp}(\mathbf{R}_a)\exp[if\beta'_0(\omega-\omega_0)z_a]$
and $\beta=\beta_0+\beta'_0(\omega-\omega_0)$ 
to calculate the integral with respect to $\omega$ in expression (\ref{v31}).
In addition, we extend the lower bound of the frequency integration to $-\infty$. 
This procedure artificially restores the effects of the missing counter-rotating terms in the Hamiltonian \cite{Loudon}. As a result, we obtain
\begin{eqnarray}\label{v32}
A_{\mathcal{N}fp}(z,t)&=&A_{\mathcal{N}fp}^{(\mathrm{in})}(z,t)
+\sqrt{2\pi}G^*_{\omega_0\mathcal{N}fp}
e^{-i(\omega_0t-f\beta_0z)}
\nonumber\\&&\mbox{}\times
\sigma(t-|z-z_a|/v_g)\Theta[f(z-z_a)]
\nonumber\\&&\mbox{}\times
\Theta(t-|z-z_a|/v_g -t_0),
\end{eqnarray}
where
\begin{equation}\label{v33}
A_{\mathcal{N}fp}^{(\mathrm{in})}(z,t)=\frac{1}{\sqrt{2\pi}}\int_0^{\infty}d\omega\, a_{\omega\mathcal{N} fp}(t_0)
e^{-i(\omega t-f\beta z)}
\end{equation}
is the injected field. In Eq.~(\ref{v32}), the coupling coefficient $G_{\omega_0\mathcal{N}fp}$ is evaluated at the atomic transition frequency $\omega_0$ and the atomic position $\mathbf{R}_a$. The notation $v_g=1/(d\beta/d\omega)$ stands for the group velocity and is evaluated at the atomic transition frequency $\omega_0$. The notation $\Theta(x)$ stands for the Heaviside step function, equal to zero for negative argument and one for positive argument. 

We study the case where the input guided pulse is prepared in a Fock state of $N$ photons, propagates in a direction $f_L=\pm$ along the fiber axis, and has a pulse shape $F_t$. The flux of transmitted photons at a position $z$ satisfying the condition $f_L(z-z_a)>0$ is given by $I_T(z,t)=I_{f=f_L}(z,t)$. When we insert Eq.~\eqref{v32} into Eq.~\eqref{v30} and
take $f=f_L$ and $f_L(z-z_a)>0$, we obtain
\begin{eqnarray}\label{v34}
\lefteqn{I_{T}(z,t)=N|F_{t-f_Lz/v_{g_L}}|^2}
\nonumber\\&&\mbox{}
+\sum_{\mathcal{N}p}\gamma_{\mathcal{N}f_Lp}\langle\sigma_{ee}(t-|z-z_a|/v_g)\rangle_{NN}
\nonumber\\&&\mbox{}
+\sqrt{2\pi N}G_{L}F_{t-f_Lz/v_{g_L}}
\langle\sigma^\dagger(t-|z-z_a|/v_{g_L})\rangle_{N,N-1}
\nonumber\\&&\mbox{}
+\sqrt{2\pi N}G^*_{L}F^*_{t-f_Lz/v_{g_L}}
\langle\sigma(t-|z-z_a|/v_{g_L})\rangle_{N-1,N},\qquad
\end{eqnarray}
where
$\gamma_{\mathcal{N}fp}=2\pi |G_{\omega_0\mathcal{N}fp}|^2$
is the rate of spontaneous emission into the guided mode $\mathcal{N}fp$. 

Meanwhile, the flux of reflected photons
at a position $z$ satisfying the condition $f_L(z-z_a)<0$ is given by $I_R(z,t)=I_{f=-f_L}(z,t)$.
When we insert Eq.~\eqref{v32} into Eq.~\eqref{v30} and
take $f=-f_L$ and $f_L(z-z_a)<0$, we obtain 
\begin{eqnarray}\label{v35}
 I_R(z,t)&=&\sum_{\mathcal{N}p}\gamma_{\mathcal{N},-f_L,p}
 \langle\sigma_{ee}(t-|z-z_a|/v_g)\rangle_{NN}.
\end{eqnarray}

Without loss of generality, we assume that the atom is located at a point with the axial coordinate $z_a=0$. In addition, we assume that, in Eqs.~(\ref{v34}) and (\ref{v35}), the group delay $|z|/v_g$ for all guided modes $\mathcal{N}fp$ is small compared to the characteristic pulse duration $T$.
Then, Eqs.~(\ref{v34}) and (\ref{v35}) reduce to
\begin{eqnarray}\label{v36}
I_T&=&N|F_t|^2+\gamma_{g}^{(\mathrm{fw})}\langle\sigma_{ee}\rangle_{NN}
\nonumber\\&&\mbox{}
+\sqrt{2\pi N}(G_{L} F_t 
\langle\sigma^\dagger\rangle_{N,N-1} 
+G_{L}^{*} F_t^* \langle\sigma\rangle_{N-1,N})\qquad
\end{eqnarray}
and
\begin{equation}\label{v37}
I_{R}=\gamma_{g}^{(\mathrm{bw})}\langle\sigma_{ee}\rangle_{NN},
\end{equation}
where $\gamma_{g}^{(\mathrm{fw})}=\gamma_g^{(f_L)}$ and $\gamma_{g}^{(\mathrm{bw})}=\gamma_g^{(-f_L)}$ are the rates of spontaneous emission into guided modes in the forward direction $f=f_L$ and the backward direction $f=-f_L$, respectively. Here, $\gamma_g^{(f)}=\sum_{\mathcal{N}p}\gamma_{\mathcal{N}fp}$ is the rate of spontaneous emission into guided modes with the propagation direction $f$. 

The expression on the right-hand side of Eq.~(\ref{v36}) has three terms. The first term, $N|F_t|^2$, is the flux of the incident field. The second term, $\gamma_{g}^{(\mathrm{fw})}\langle\sigma_{ee}\rangle_{NN}$, is the rate of scattering into guided modes in the forward direction $f=f_L$. The last term, proportional to $\sqrt{N}(G_{L} F_t 
\langle\sigma^\dagger\rangle_{N,N-1}+\mathrm{c.c.})$, describes the effect of the interference between the incident and forward scattered fields. Meanwhile,
the expression on the right-hand side of Eq.~(\ref{v37}) is the rate of scattering into guided modes in the backward direction $f=-f_L$. 
According to Eq.~(\ref{v37}), the photon reflection flux $I_R$ and the atomic excitation probability $\langle\sigma_{ee}\rangle_{NN}$ are proportional to each other. Consequently, the time dependencies of $I_R$ and $\langle\sigma_{ee}\rangle_{NN}$ have the same shape. 

We introduce the notation $I_{\mathrm{rad}}=\gamma_r \langle\sigma_{ee}\rangle_{NN}$ for the rate of scattering into radiation modes, where $\gamma_r$ is the rate of spontaneous emission into radiation modes. We find the relation $I_T+I_R+I_{\mathrm{rad}}+\langle\dot{\sigma}_{ee}\rangle_{NN}=N|F_t|^2$, 
in agreement with the energy conservation law. 

We introduce the notations $P_T=\int_{t_0}^{\infty}I_T(t)dt$ and
$P_R=\int_{t_0}^{\infty}I_R(t)dt$ for the mean numbers of transmitted and reflected photons, respectively. We also introduce the notation $P_{\mathrm{rad}}=\int_{t_0}^{\infty}I_{\mathrm{rad}}(t)dt$ for the mean number of photons scattered into radiation modes. We find
$P_T+P_R+P_{\mathrm{rad}}=N$.
The extinction of the pulse is $P_{\mathrm{ext}}=N-P_T=P_R+P_{\mathrm{rad}}$.

In the case of single-photon pulses ($N=1$), we can rewrite Eqs.~(\ref{v36}) and (\ref{v37}) in the form
\begin{equation}\label{v36a}
I_T=|F_t|^2+\gamma_{g}^{(\mathrm{fw})}P+\sqrt{2\pi}(G_{L} F_t Q^*+G_{L}^{*} F_t^* Q)
\end{equation}
and
\begin{equation}\label{v37c}
I_{R}=\gamma_{g}^{(\mathrm{bw})}P.
\end{equation}
In addition, we find $I_T+I_R+I_{\mathrm{rad}}+\dot{P}=|F_t|^2$
and $P_T+P_R+P_{\mathrm{rad}}=1$. 
Equations (\ref{v36a}) and (\ref{v37c}) are in agreement with the results of Ref.~\cite{Domokos2002} for single-photon light pulses. With the help of the relation $P=|Q|^2$, valid for the case of single-photon pulses, we can rewrite Eq.~(\ref{v36a}) as $I_T=(\gamma_{g}^{(\mathrm{fw})}-\gamma_L)P+|F_t+\sqrt{2\pi}G_L^*Q|^2$. 
In the particular case where $\gamma_{g}^{(\mathrm{fw})}=\gamma_L$,
we obtain $I_T=|F_t+\sqrt{2\pi}G_L^*Q|^2$, in agreement with the results of Ref.~\cite{Kurtsiefer2016}. 
 
In the case of single-photon pulses, $P_T$ and $P_R$ are the probabilities of transmission and reflection, respectively, $P_{\mathrm{rad}}$ is the probability of scattering into radiation modes, and
$P_{\mathrm{ext}}=1-P_T=P_R+P_{\mathrm{rad}}$ is the extinction probability.
When we integrate the atomic excitation probability $P(t)$ over the time $t$ for the whole interaction process, we obtain the quantity
$\tau_{e}=\int_{t_0}^{\infty} P(t) dt$,
which can be called the effective excitation time of the atom. 
With the help of Eqs.~(\ref{v27})--(\ref{v29}), we can show that single-photon rising and decaying exponential pulses with the same pulse duration $T$ produce the same effective excitation time
\begin{equation}\label{v37b2}
\tau_e=4T\frac{\gamma_L}{\gamma}\frac{1+\gamma T}{(1+\gamma T)^2+4\Delta^2 T^2}.
\end{equation}
Hence, the reflection probability $P_R=\gamma_{g}^{(\mathrm{bw})}\tau_{e}$, the probability of emission into radiation modes $P_{\mathrm{rad}}=\gamma_r\tau_{e}$, the extinction probability $P_{\mathrm{ext}}=(\gamma_{g}^{(\mathrm{bw})}+\gamma_r)\tau_{e}$, and the transmission probability $P_T=1-(\gamma_{g}^{(\mathrm{bw})}+\gamma_r)\tau_{e}$ 
do not depend on whether the pulse is exponentially rising or decaying \cite{Kurtsiefer2016}. 
	 
We note that, in the case where the injected pulse is prepared in a coherent state $\alpha$,
we have 
\begin{eqnarray}\label{v38}
	I_T&=&|\alpha|^2|F_t|^2+\gamma_g^{(\mathrm{fw})}\langle\sigma_{ee}\rangle
	+\sqrt{2\pi} (\alpha G_{L} F_t \langle\sigma^\dagger\rangle 
	\nonumber\\&&\mbox{}
		+\alpha^* G_{L}^{*} F_t^* \langle\sigma\rangle)
\end{eqnarray}
and
\begin{equation}\label{v39}
I_{R}=\gamma_{g}^{(\mathrm{bw})}\langle\sigma_{ee}\rangle.
\end{equation}

It is worth mentioning that, by using appropriate expressions for the coupling coefficient $G_L$, we can apply Eqs.~(\ref{v36}) and (\ref{v37}) to not only quasicircularly polarized modes but also quasilinearly polarized modes. 

\subsection{Chiral coupling between an atom and a quasilinearly polarized hybrid guided field}

In this subsection, we study the dependence of the interaction between the atom and the guided probe pulse on the propagation direction of the pulse.

We assume that the probe pulse is prepared in a quasilinearly polarized hybrid guided mode $\mu_L$ of the nanofiber.
Quasilinearly polarized hybrid modes are linear superpositions of counterclockwise and clockwise quasicircularly polarized hybrid modes. The amplitude of the guided field in a quasilinearly polarized hybrid mode can be written in the form \cite{highorder}
\begin{eqnarray}\label{c12}
\mathbf{e}^{(\mu_L)}&=&\sqrt2[e_r\cos (l\varphi-\varphi_{\mathrm{pol}})\,\hat{\mathbf{r}}
+i e_\varphi\sin (l\varphi-\varphi_{\mathrm{pol}})\,\hat{\boldsymbol{\varphi}}
\nonumber\\&&\mbox{}\times
+f_Le_z\cos (l\varphi-\varphi_{\mathrm{pol}})\,\hat{\mathbf{z}}],
\end{eqnarray}
where $e_r$, $e_\varphi$, and $e_z$ are the cylindrical components of 
the profile function of the corresponding quasicircularly polarized hybrid guided modes and are evaluated at the frequency $\omega=\omega_0$.
The phase angle $\varphi_{\mathrm{pol}}$ determines the orientation of the symmetry axes of the mode profile in the fiber transverse plane. In particular, the specific values $\varphi_{\mathrm{pol}}=0$ and $\pi/2$ define two orthogonal polarization profiles, called even and odd, respectively.
We again use the notation $\mathbf{R}=(r,\varphi,z)$ for the position of the atom. 

The coupling coefficient $G_L$ for the atom and the quasilinearly polarized hybrid guided field is given as
\begin{equation}\label{v2a}
G_L=\sqrt{\frac{\omega_0\beta_L'}{4\pi\epsilon_0\hbar}}\;
(\mathbf{d}\cdot\mathbf{e}^{(\mu_L)})e^{if_L\beta_L z},
\end{equation}
where $\beta_L$ and $\beta_L'$ are evaluated at the frequency $\omega=\omega_0$.
We assume that the atom is located on the positive side of the $x$ axis, that is, $\varphi=0$.
When we insert Eq.~(\ref{c12}) into Eq.~(\ref{v2a}), we obtain
\begin{equation}\label{v2b1}
|G_L(\varphi_{\mathrm{pol}}=0)|=\sqrt{\frac{\omega_0\beta_L'}{2\pi\epsilon_0\hbar}}\; |d_xe_r+f_Ld_ze_z|
\end{equation}
and
\begin{equation}\label{v2b2}
|G_L(\varphi_{\mathrm{pol}}=\pi/2)|=\sqrt{\frac{\omega_0\beta_L'}{2\pi\epsilon_0\hbar}}\;
|d_y e_\varphi|.
\end{equation}
Here, $d_x=d_r$, $d_y=d_\varphi$, and $d_z$ are the components of the dipole matrix element vector $\mathbf{d}$ in the Cartesian and cylindrical coordinate systems.

According to Eq.~(\ref{v2b2}), the absolute value $|G_L|$ of the coupling coefficient $G_L$ for the quasilinearly polarized guided mode of the odd type (with the polarization angle $\varphi_{\mathrm{pol}}=\pi/2$) does not depends on the propagation direction $f_L$. The reason is that the polarization of the field at the position of the atom is linear. 

Meanwhile, Eq.~(\ref{v2b1}) shows that the absolute value $|G_L|$ of the coupling coefficient $G_L$ for the quasilinearly polarized guided mode of the even type (with the polarization angle $\varphi_{\mathrm{pol}}=0$) depends on the field propagation direction $f_L$ if 
\begin{equation}\label{v2c}
\mathrm{Re}\,(d_xd_z^*e_re_z^*)\not=0.
\end{equation}
It is known that both the radial component $e_r$ and the axial component $e_z$ of the mode function of quasicircularly polarized hybrid modes are nonzero and their relative phase is $\pi/2$ \cite{fiber books,highorder}. Hence, condition (\ref{v2c}) reduces to the condition
\begin{equation}\label{v2d}
\mathrm{Im}\,(d_xd_z^*)\not=0
\end{equation} 
for the atomic dipole. This condition means that the atom has a dipole rotating in the meridional plane $zx$, that is, the atom is chiral. The ellipticity vector of the dipole of this atom overlaps with the ellipticity vector of the quasilinearly polarized field mode of the even type \cite{Fam2014,Petersen2014,Mitsch14b,Lodahl2017,sponhigh}.
The directional dependence of the absolute value of the coupling coefficient $G_L$ leads to the directional dependence of the coupling parameter $\gamma_L=2\pi |G_L|^2$ and, hence, to the directional dependence of the atomic excitation probability $P$ [see Eq.~(\ref{v25a})].

Similar to the probe-atom coupling parameter $\gamma_L$, the rate $\gamma_{g}^{(f)}$ of spontaneous emission
into guided modes in the direction $f$ under condition (\ref{v2d}) is asymmetric with respect to the opposite directions $f=+$ and $f=-$ \cite{Fam2014,Petersen2014,Mitsch14b,Lodahl2017,sponhigh}. 
The directional dependencies of $\gamma_L$ and $\gamma_{g}^{(f)}$ are the signatures of spin-orbit coupling of light carrying transverse spin angular momentum \cite{Zeldovich,Bliokh review,Bliokh review2015,Bliokh2014,Bliokh2015,Lodahl2017,Banzer review2015}. They are due to the existence of a nonzero longitudinal component of the field in the presence of the nanofiber. This component oscillates in phase quadrature with respect to the radial transverse component and, hence, makes the field chiral.
The directional dependencies of $\gamma_L$ and $\gamma_{g}^{(f)}$
determine the directional dependencies of the transmission and reflection fluxes and the corresponding transmission and reflection probabilities.

\section{Numerical results}
\label{sec:numerical}

In this section, we present the results of numerical calculations for the interaction between the atom and a quantized light pulse in a guided mode of the nanofiber.

We use the atomic transition wavelength $\lambda_0=852$ nm and
the natural linewidth $\gamma_0/2\pi=5.2$ MHz, which correspond to the transitions in the $D_2$ line of atomic cesium. 
The atomic dipole matrix element $d$ is calculated from the formula $\gamma_0=d^2\omega_0^3/3\pi\epsilon_0\hbar c^3$ for the natural linewidth of a two-level atom \cite{Loudon,Scully}.

In order to maximize the coupling efficiency between the guided probe field and the atom, we use
a single-mode nanofiber.
We assume that the fiber radius is $a=200$ nm, and the refractive indices of the fiber and the vacuum cladding are $n_1=1.45$ and $n_2=1$, respectively. 
This thin fiber supports only the fundamental mode HE$_{11}$ at the wavelength $\lambda_0$ of the atom considered. The quasilinearly polarized
HE$_{11}$ modes with the polarization angles $\varphi_{\mathrm{pol}}=0$ and $\pi/2$ of the nanofiber are called
$x$- and $y$-polarized guided modes, respectively.
We assume that
the injected field is prepared in the $x$-polarized guided mode.
We note that the spatial intensity distribution of the injected field is maximal on the $x$ axis,
where the atom is positioned. In order to get a chiral effect in the interaction between the atom and the probe guided light field, we consider the case where 
the atomic dipole rotates in the meridional plane containing the atomic position. In this case, the dipole matrix element vector $\mathbf{d}$ is a complex vector in the $zx$ plane.
To be concrete, we take $\mathbf{d}=d(i\hat{\mathbf{x}}-\hat{\mathbf{z}})/\sqrt2$. This matrix element corresponds to a $\sigma_+$-type transition between the magnetic levels of an alkali-metal atom that are specified with the use of the axis $y$ as the quantization axis. Since condition (\ref{v2d}) is satisfied,
the absolute value of the coupling coefficient for the atom and the $x$-polarized fundamental mode depends on the field propagation direction [see Eq.~(\ref{v2b1})]. Meanwhile, since $d_y=0$, the atom does not interact with the $y$-polarized fundamental mode [see Eq.~(\ref{v2b2})]. Hence, we have $\gamma_L=\gamma_g^{(f_L)}=\gamma_g^{(\mathrm{fw})}$, that is, the probe-atom coupling parameter $\gamma_L$ is equal to the rate $\gamma_g^{(f_L)}$ of spontaneous emission into guided modes in the forward direction $f_L$. 

\begin{figure}[tbh]
	\begin{center}
		\includegraphics{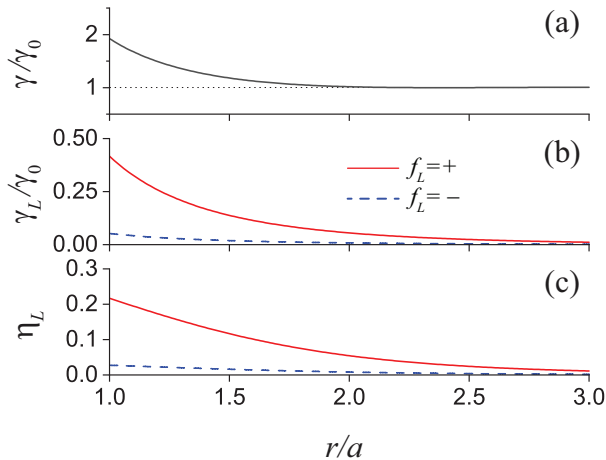}
	\end{center}
	\caption{Radial dependencies of the total spontaneous emission rate $\gamma$ (a), the probe-atom coupling parameter $\gamma_L=2\pi |G_L|^2$ (b), and the coupling efficiency $\eta_L=\gamma_L/\gamma$ (c). The fiber radius is $a=200$ nm. The refractive indices of the silica core and the vacuum cladding are $n_1=1.45$ and $n_2=1$, respectively. The atom is positioned on the $x$ axis. The input field mode $\mu_L$ is quasilinearly polarized in the $x$ direction and propagates along the fiber axis in the direction $f_L=+$ (solid red lines) or $-$ (dashed blue lines). The dipole matrix element vector of the atom is $\mathbf{d}=d(i\hat{\mathbf{x}}-\hat{\mathbf{z}})/\sqrt2$. The dipole magnitude corresponds to the natural linewidth $\gamma_0/2\pi=5.2$ MHz of the $D_2$ line of atomic cesium with the transition wavelength $\lambda_0=852$ nm.
	}
	\label{fig2}
\end{figure}

We calculate the total spontaneous emission rate $\gamma$, 
the probe-atom coupling parameter $\gamma_L=2\pi |G_L|^2$, and the coupling efficiency $\eta_L=\gamma_L/\gamma$.
We plot in Fig.~\ref{fig2} the radial dependencies of these characteristics.
We observe from the figure that $\gamma$, $\gamma_L$, and $\eta_L$ reduce quickly with increasing distance from the atom to the fiber surface. Figures \ref{fig2}(b) and \ref{fig2}(c) show that the values of the coupling parameter $\gamma_L$ and the coupling efficiency $\eta_L$ for the probe field with the propagation direction $f_L=+$ (solid red lines) are larger than those for the probe field with the propagation direction $f_L=-$ (dashed blue lines). It follows from the dependence of $\gamma_L$ on $f_L$ and the relation $\gamma_L=\gamma_g^{(\mathrm{fw})}$ that the rates $\gamma_g^{(\mathrm{fw})}$ and $\gamma_g^{(\mathrm{bw})}$ of spontaneous emission into guided modes in the forward and backward directions also depend on $f_L$. We show below that
the directional dependencies of the coupling parameter $\gamma_L$ and the rates $\gamma_g^{(\mathrm{fw})}$ and $\gamma_g^{(\mathrm{bw})}$ lead to the directional dependencies of the atomic excitation probability, the photon transmission flux, and the photon transmission probability.

\subsection{Atomic excitation probability}

\begin{figure}[tbh]
\begin{center}
 \includegraphics{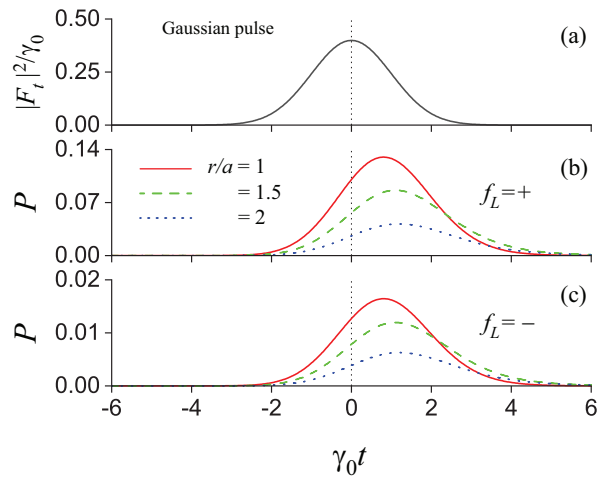}
 \end{center}
\caption{Excitation of the atom by a single-photon Gaussian light pulse in the $x$-polarized fundamental mode HE$_{11}$. (a) Temporal pulse profile function $|F_t|^2$. (b),(c) 
	Time dependence of the atomic excitation probability $P$ of the atom
	interacting with the pulse with the propagation direction $f_L=+$ (b) or $f_L=-$ (c). The radial position of the atom is $r/a=1$ (solid red lines), 1.5 (dashed green lines), and 2 (dotted blue lines). 
	The quantized pulse is at exact resonance with the atom. The characteristic pulse length is $T=1/\gamma_0\simeq 30$ ns. Other parameters are as for Fig.~\ref{fig2}.
	The vertical dotted black line indicates the pulse peak time $t=0$. 
}
\label{fig3}
\end{figure}

We use Eqs.~(\ref{v23}) or the analytical expressions (\ref{v25a})--(\ref{v29})
to calculate the internal state of the atom interacting with a single-photon guided light pulse.
We plot in Fig.~\ref{fig3} the time dependence of the atomic excitation probability $P$ for the case of a single-photon Gaussian guided light pulse. We observe from the solid red curve of Fig.~\ref{fig3}(b) that, for an atom at the fiber surface, the excitation probability $P$ can be as large as $\approx 0.13$ even though the incident guided light pulse has just a single photon. Comparison between different curves of Fig.~\ref{fig3}(b) as well as Fig.~\ref{fig3}(c) shows that the peak value of the excitation probability decreases with increasing distance from the atom to the fiber surface. This behavior is a consequence of the evanescent-wave nature of the guided field. We observe that the arrival of the peak is delayed by a significant amount of time, which is comparable to the free-space lifetime $\tau_0=1/\gamma_0$ of the atom. More importantly, comparison between Figs.~\ref{fig3}(b) and \ref{fig3}(c) shows that the excitation probability $P$ strongly depends on the propagation direction $f_L$ of the pulse. The directional dependence of $P$ is a chiral effect and is a consequence of spin-orbit coupling of guided light carrying transverse spin angular momentum \cite{Zeldovich,Bliokh review,Bliokh review2015,Bliokh2014,Bliokh2015,Lodahl2017,Banzer review2015}.

\begin{figure}[tbh]
	\begin{center}
		\includegraphics{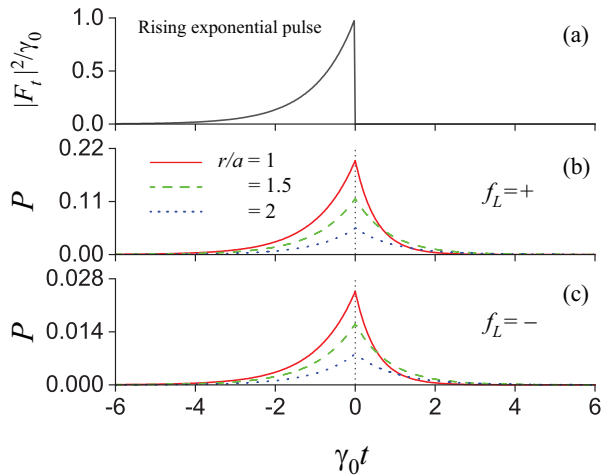}
	\end{center}
	\caption{Excitation of the atom by a single-photon rising exponential light pulse in the $x$-polarized fundamental mode HE$_{11}$. (a) Temporal pulse profile function $|F_t|^2$. (b),(c) Time dependence of the atomic excitation probability $P$ of the atom interacting with the pulse with the propagation direction $f_L=+$ (b) or $f_L=-$ (c). Other parameters are as for Fig.~\ref{fig3}. 
	}
	\label{fig4}
\end{figure}

\begin{figure}[tbh]
	\begin{center}
		\includegraphics{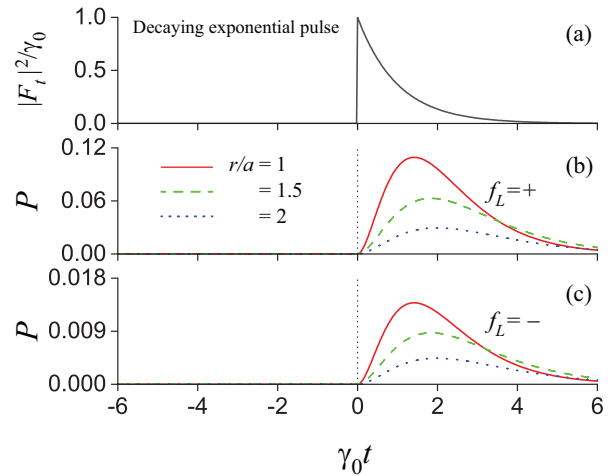}
	\end{center}
	\caption{Excitation of the atom by a single-photon decaying exponential light pulse in the $x$-polarized fundamental mode HE$_{11}$. (a) Temporal pulse profile function $|F_t|^2$. (b),(c) Time dependence of the atomic excitation probability $P$ of the atom interacting with the pulse with the propagation direction $f_L=+$ (b) or $f_L=-$ (c). Other parameters are as for Fig.~\ref{fig3}. 
	}
	\label{fig5}
\end{figure}

We plot in Figs.~\ref{fig4} and \ref{fig5} the time dependencies of the atomic excitation probability $P$ for single-photon rising and decaying exponential pulses. We observe that the excitation probability of the atom substantially depends on the pulse shape \cite{Wang2011,Wang2012,GB,Kurtsiefer2016}. Comparison between Figs.~\ref{fig3}--\ref{fig5} shows that the rising exponential pulse shape is more favorable to excite atoms than the other pulse shapes. The magnitude of $P$ can be as high as $\approx 0.2$, achieved for a rising exponential pulse interacting with an atom at the fiber surface [see the solid red line in Fig.~\ref{fig4}(b)]. Comparison between Figs.~\ref{fig4}(b) and \ref{fig4}(c) and between Figs.~\ref{fig5}(b) and \ref{fig5}(c) confirms that, like in the case of Fig.~\ref{fig3}, the excitation probability $P$ in the cases of Figs.~\ref{fig4} and \ref{fig5} strongly depends on the propagation direction $f_L$ of the pulse. We observe again that the peak value of $P$ decreases with increasing distance from the atom to the fiber surface.

\begin{figure}[tbh]
	\begin{center}
		\includegraphics{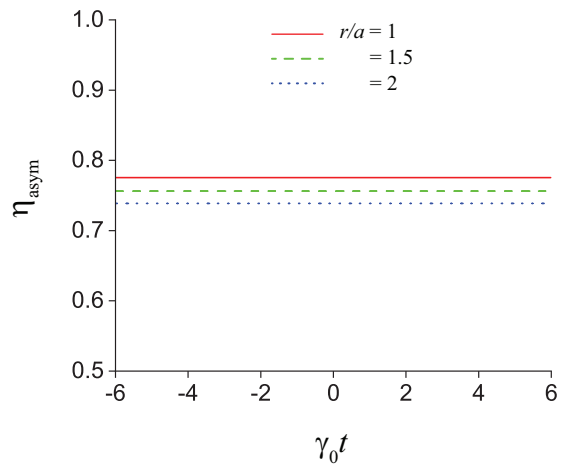}
	\end{center}
	\caption{Time dependence of the asymmetry parameter $\eta_{\mathrm{asym}}$ for the excitation probability of the atom interacting with a single-photon guided light pulse. The shape of the pulse is arbitrary. Other parameters are as for Fig.~\ref{fig3}. 		
	}
	\label{fig6}
\end{figure}

The relative difference between the excitation probabilities $P_{\pm}=P$ for the opposite propagation directions $f_L=\pm$ can be characterized by the asymmetry parameter $\eta_{\mathrm{asym}}=(P_+-P_-)/(P_++P_-)$. 
It follows from Eq.~(\ref{v25a}) that $\eta_{\mathrm{asym}}=(\gamma_L^{(+)}-\gamma_L^{(-)})/(\gamma_L^{(+)}+\gamma_L^{(-)})$,
where $\gamma_L^{(\pm)}=\gamma_L$ for $f_L=\pm$. It is clear that the asymmetry parameter 
$\eta_{\mathrm{asym}}$ does not vary in time and does not depend on the pulse shape. 
We observe these features in Fig.~\ref{fig6}, where the asymmetry parameter $\eta_{\mathrm{asym}}$ is plotted as a function of time for single-photon light pulses of arbitrary shape.

\subsection{Photon reflection and transmission fluxes}

\begin{figure}[tbh]
	\begin{center}
		\includegraphics{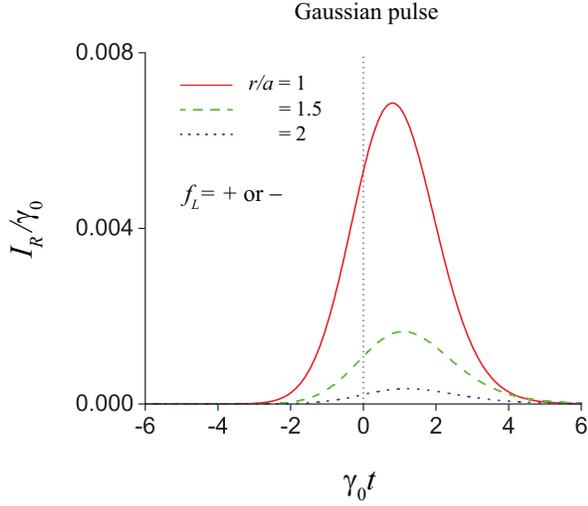}
	\end{center}
	\caption{Time dependence of the photon reflection flux $I_R$ for a single-photon Gaussian guided light pulse. The propagation direction of the pulse is $f_L=+$ or $-$. Other parameters are as for Fig.~\ref{fig3}.		
	}
	\label{fig7}
\end{figure}

\begin{figure}[tbh]
	\begin{center}
		\includegraphics{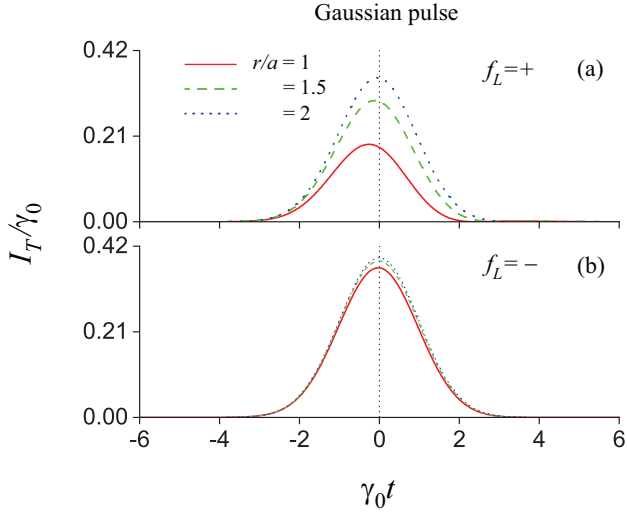}
	\end{center}
	\caption{Time dependence of the photon transmission flux $I_T$ for a single-photon Gaussian guided light pulse. The propagation direction of the pulse is $f_L=+$ (a) or $-$ (b). Other parameters are as for Fig.~\ref{fig3}.		
	}
	\label{fig8}
\end{figure}

We use Eqs.~(\ref{v36a}) and (\ref{v37c}) to calculate the photon reflection flux $I_R$ and the photon transmission flux $I_T$. We plot in Figs.~\ref{fig7} and \ref{fig8} the results of calculations for the time dependencies of the fluxes $I_R$ and $I_T$ for the atom interacting with a single-photon Gaussian pulse. Comparison between Figs.~\ref{fig3} and \ref{fig7} shows that the time dependencies of the atomic excitation probability $P$ and the photon reflection flux $I_R$ have the same shape, in agreement with Eq.~(\ref{v37c}). Like the peak of $P$ in Fig.~\ref{fig3}, the peak of $I_R$ in Fig.~\ref{fig7} is delayed by a significant amount of time.

The numerical results presented in Fig.~\ref{fig7} show that, unlike the atomic excitation probability $P$, the photon reflection flux $I_R$ does not depend on the propagation direction $f_L$ of the probe pulse. This feature is a consequence of the fact that the reflection involves two processes, namely the atomic excitation by the pulse propagating in one direction [see Eq.~(\ref{v25a})] and the subsequent photon re-emission into the guided modes propagating in the opposite direction [see Eq.~(\ref{v37c})]. Due to this fact, the dependence of $I_R$ on $f_L$ is contained in the proportionality factor $\gamma_g^{(\mathrm{bw})}\gamma_L$ [see Eqs.~(\ref{v25a}) and (\ref{v37c})].
For the considered atomic dipole and guided probe pulse, we have $\gamma_L=\gamma_g^{(\mathrm{fw})}$. Therefore, the proportionality factor is
$\gamma_g^{(\mathrm{bw})}\gamma_L=\gamma_g^{(\mathrm{bw})}\gamma_g^{(\mathrm{fw})}=\gamma_g^{(+)}\gamma_g^{(-)}$. It is clear that this factor does not depend on $f_L$ and hence neither does the reflection flux $I_R$.

Comparison between Figs.~\ref{fig8}(a) and \ref{fig8}(b) shows that the photon transmission flux $I_T$ depends on the field propagation direction $f_L$. In the case of the solid red curve in Fig.~\ref{fig8}(a), where $f_L=+$ and $r/a=1$, we observe a significant advance (negative delay) of the time for the arrival of the peak of the pulse. This advance is related to the anomalous dispersion of the susceptibility of resonant two-level atoms \cite{Scully}. 

\begin{figure}[tbh]
	\begin{center}
		\includegraphics{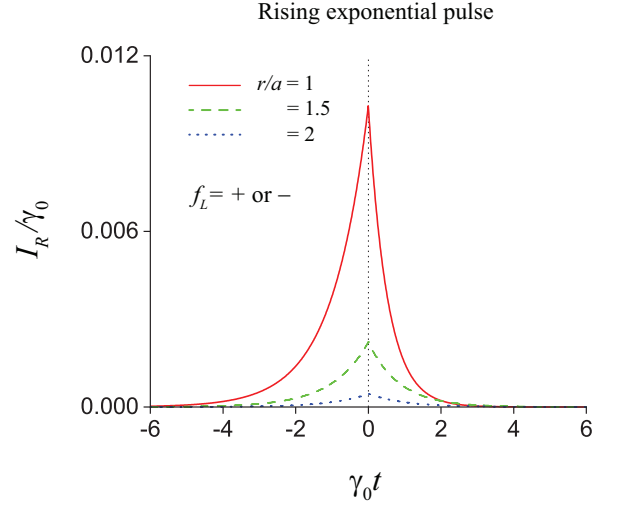}
	\end{center}
	\caption{Time dependence of the photon reflection flux $I_R$ for a single-photon rising exponential guided light pulse. The propagation direction of the pulse is $f_L=+$ or $-$. Other parameters are as for Figs.~\ref{fig3} and \ref{fig4}.		
	}
	\label{fig9}
\end{figure}

\begin{figure}[tbh]
	\begin{center}
		\includegraphics{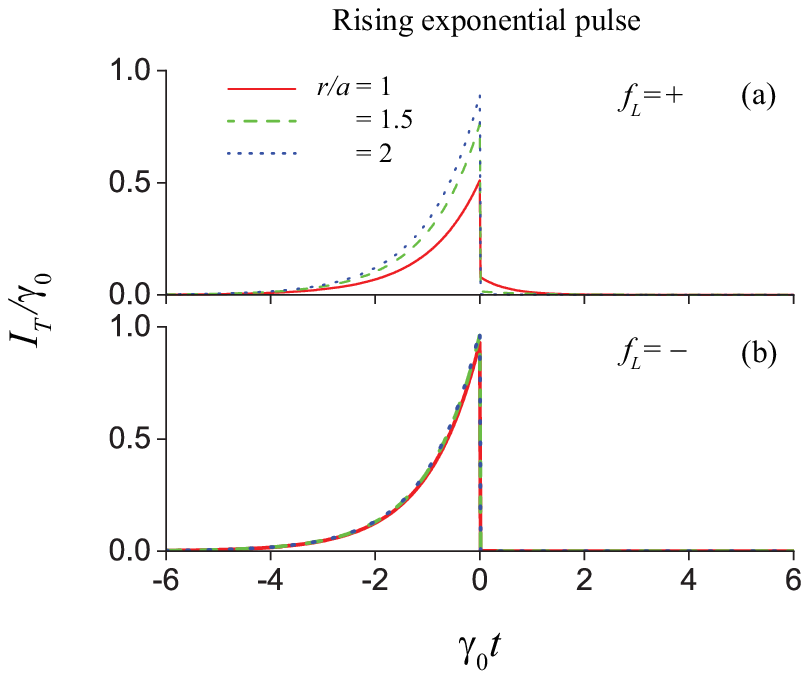}
	\end{center}
	\caption{Time dependence of the photon transmission flux $I_T$ for a single-photon rising exponential guided light pulse. The propagation direction of the pulse is $f_L=+$ (a) or $-$ (b). Other parameters are as for Figs.~\ref{fig3} and \ref{fig4}.		
	}
	\label{fig10}
\end{figure}

\begin{figure}[tbh]
	\begin{center}
		\includegraphics{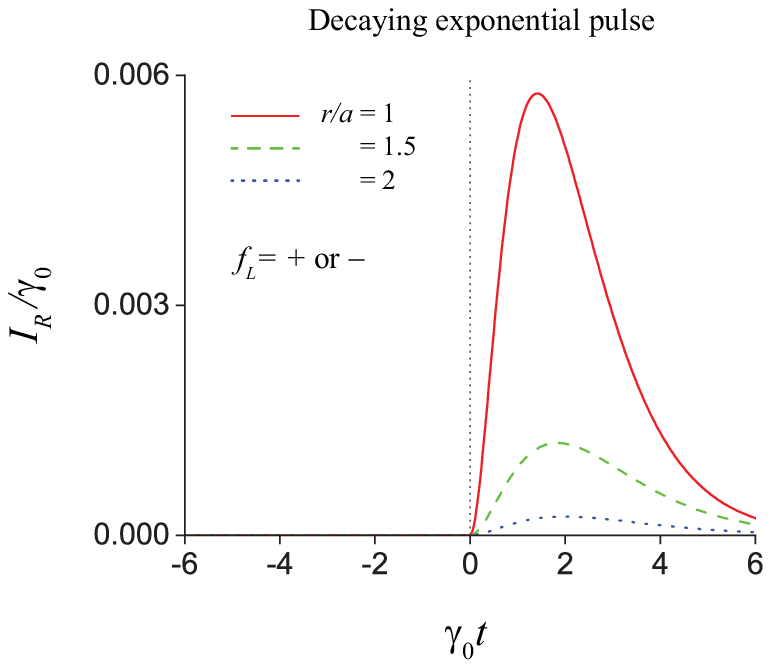}
	\end{center}
	\caption{Time dependence of the photon reflection flux $I_R$ for a single-photon decaying exponential guided light pulse. The propagation direction of the pulse is $f_L=+$ or $-$. Other parameters are as for Figs.~\ref{fig3} and \ref{fig5}.		
	}
	\label{fig11}
\end{figure}

\begin{figure}[tbh]
	\begin{center}
		\includegraphics{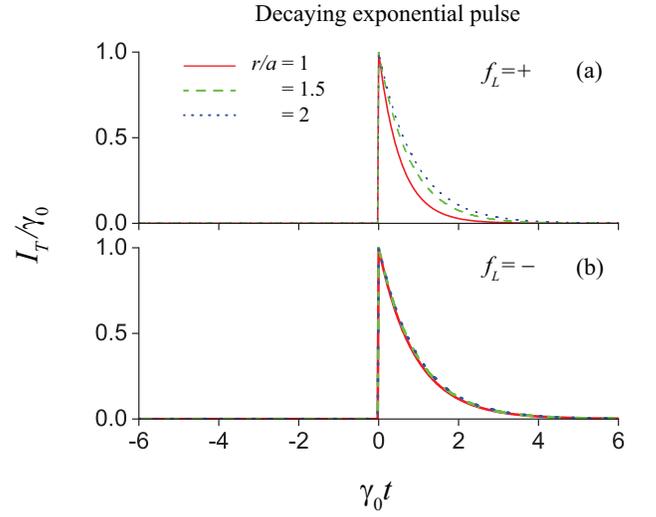}
	\end{center}
	\caption{Time dependence of the photon transmission flux $I_T$ for a single-photon decaying exponential guided light pulse. The propagation direction of the pulse is $f_L=+$ (a) or $-$ (b). Other parameters are as for Figs.~\ref{fig3} and \ref{fig5}.		
	}
	\label{fig12}
\end{figure}

We plot in Figs.~\ref{fig9}--\ref{fig12} the time dependencies of the photon reflection and transmission fluxes $I_R$ and $I_T$ for single-photon rising and decaying exponential pulses. We observe that the temporal shapes of the reflection and transmission fluxes substantially depend on the pulse shape. Like in the case of Gaussian pulses, we observe in the cases of rising and decaying exponential pulses that the reflection flux $I_R$ does not depend on the 
pulse propagation direction $f_L$, while the transmission flux $I_T$ depends on $f_L$.

\subsection{Photon reflection and transmission probabilities}

\begin{figure}[tbh]
	\begin{center}
		\includegraphics{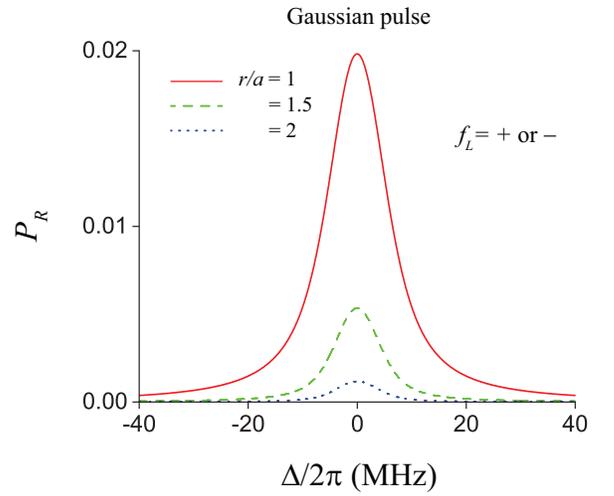}
	\end{center}
	\caption{Dependence of the reflection probability $P_R$ on the field detuning $\Delta$ of a single-photon Gaussian guided light pulse. The propagation direction of the pulse is $f_L=+$ or $-$. Other parameters are as for Fig.~\ref{fig3}.		
	}
	\label{fig13}
\end{figure}

\begin{figure}[tbh]
	\begin{center}
		\includegraphics{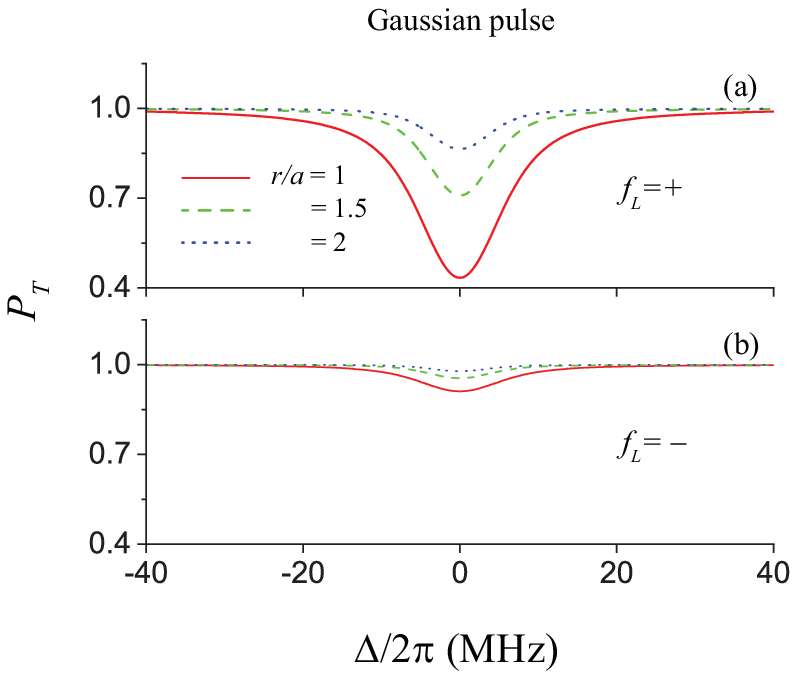}
	\end{center}
	\caption{Dependence of the transmission probability $P_T$ on the field detuning $\Delta$ of a single-photon Gaussian guided light pulse. The propagation direction of the pulse is $f_L=+$ (a) or $-$ (b). Other parameters are as for Fig.~\ref{fig3}.		
	}
	\label{fig14}
\end{figure}

\begin{figure}[tbh]
	\begin{center}
		\includegraphics{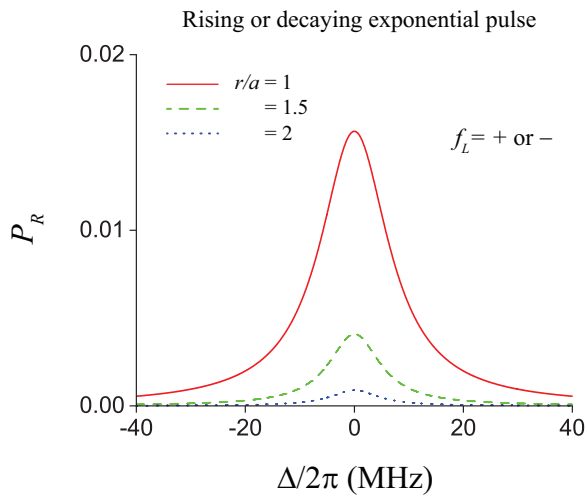}
	\end{center}
	\caption{Dependence of the reflection probability $P_R$ on the field detuning $\Delta$ of a single-photon rising or decaying exponential guided light pulse. The propagation direction of the pulse is $f_L=+$ or $-$. Other parameters are as for Figs.~\ref{fig3}--\ref{fig5}.		
	}
	\label{fig15}
\end{figure}

\begin{figure}[tbh]
	\begin{center}
		\includegraphics{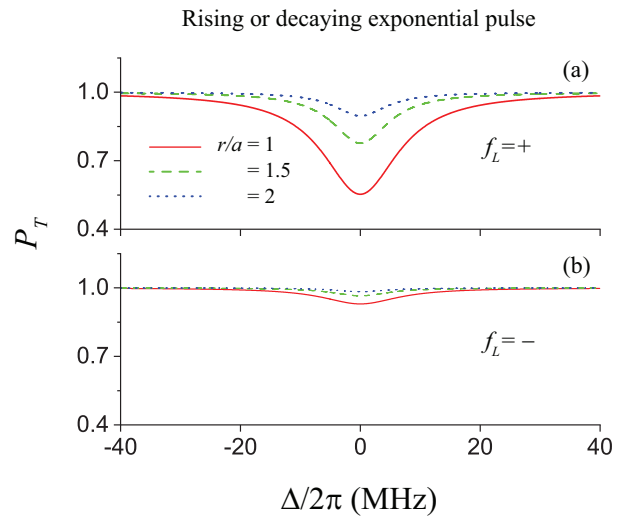}
	\end{center}
	\caption{Dependence of the transmission probability $P_T$ on the field detuning $\Delta$ of a single-photon rising or decaying exponential guided light pulse. The propagation direction of the pulse is $f_L=+$ (a) or $-$ (b). Other parameters are as for Fig.~\ref{fig3}--\ref{fig5}.		
	}
	\label{fig16}
\end{figure}

We calculate the photon reflection probability $P_R=\int_{t_0}^{\infty}I_R(t)dt$ and the photon transmission probability $P_T=\int_{t_0}^{\infty}I_T(t)dt$ by integrating the corresponding fluxes. We plot in Figs.~\ref{fig13} and \ref{fig14} the dependencies of $P_R$ and $P_T$ on the field detuning $\Delta=\omega_L-\omega_0$ for a single-photon Gaussian pulse. 
We depict in Figs.~\ref{fig15} and \ref{fig16} the corresponding results for single-photon rising and decaying exponential pulses. It is clear that the curves are symmetric with respect to $\Delta$. According to the numerical results presented in Figs.~\ref{fig13} and \ref{fig15}, the reflection probability $P_R$ has the same magnitude for pulses with the opposite propagation directions $f_L=\pm$. This feature occurs as a consequence of the fact that the photon reflection flux $I_R$ does not depend on the propagation direction $f_L$ of the pulse. Comparison between Figs.~\ref{fig14}(a) and \ref{fig14}(b) and between Figs.~\ref{fig16}(a) and \ref{fig16}(b) shows that the transmission probability $P_T$ depends on the field propagation direction $f_L$. We observe from Figs.~\ref{fig13}--\ref{fig16} that the linewidths of the curves for the frequency dependencies of $P_R$ and $P_T$ increase with decreasing distance from the atom to the fiber surface. This feature is a consequence of the dependence of the total decay rate $\gamma$ on the radial position of the atom [see Fig.~\ref{fig2}(a)].
The numerical results presented in Figs.~\ref{fig15} and \ref{fig16} show that
the probabilities $P_R$ and $P_T$ do not depend on whether the single-photon probe pulse is exponentially rising or decaying. This result is in agreement with the results of Ref.~\cite{Kurtsiefer2016} for the extinction probability $P_{\mathrm{ext}}=1-P_T$ of a single photon interacting with a single trapped atom. Comparison between the curves of Figs.~\ref{fig13}--\ref{fig16} shows that, for increasing distance from the atom to the fiber surface, the reflection probability $P_R$ decreases and the transmission probability $P_T$ increases.

\section{Summary}
\label{sec:summary}

In conclusion, we have studied the interaction between a single two-level atom and a single-photon probe pulse in a guided mode of a nanofiber. We have focused on the situation of chiral interaction where the atom has a dipole rotating in the meridional plane of the nanofiber, and the probe pulse is quasilinearly polarized along the radial direction of the position of the atom in the fiber transverse plane. We have shown that, for increasing distance from the atom to the fiber surface, 
the peak atomic excitation probability and the photon reflection probability decrease, while the photon transmission probability increases. We have found that the atomic excitation probability, the photon transmission flux, and the photon transmission probability depend on the propagation direction of the probe pulse along the fiber axis. These directional dependencies are the consequences of spin-orbit coupling of light carrying transverse spin angular momentum. We have shown that the asymmetry parameter for the atomic excitation probability does not vary in time and does not depend on the probe pulse shape. Unlike the photon transmission flux and the photon transmission probability, the reflection flux and the reflection probability do not depend on the propagation direction of the probe pulse. In the case of single-photon Gaussian pulses, we have observed a time delay of the peak of the photon reflection flux and a time advance of the peak of the photon transmission flux. We have shown that, for an arbitrary detuning, the reflection probability and the transmission probability do not depend on whether the pulse is exponentially rising or decaying. 

Our results are important, as they can be used to control and manipulate the directional dependence of the interaction between a single atom and a single-photon guided light pulse. They can be envisioned to have significant influence on ongoing and future experiments in nanofiber quantum optics. Due to the high efficiencies that can be achieved for coupling into the fiber, our scheme could be more efficient than the scheme for single-photon scattering by a single atom in free space \cite{Kurtsiefer2016}. Our scheme can also be extended to be used as a one-atom switch for single-photon routing controlled by a single photon \cite{Dayan2014}. Compared the microcavity-based system \cite{Dayan2014}, a nanofiber-based system is likely to be less efficient, though somewhat simpler in design.

\begin{acknowledgments}
This work was supported by the Okinawa Institute of Science and Technology Graduate University.
\end{acknowledgments}




\begin{thebibliography}{99}
	
\bibitem{Cirac1997}	 J. I. Cirac, P. Zoller, H. J. Kimble, and H. Mabuchi, Phys. Rev. Lett. \textbf{78}, 3221 (1997).
	 
\bibitem{Haroche1997} X. Maitre, E. Hagley, G. Nogues, C. Wunderlich, P. Goy, M. Brune, J. M. Raimond, and S. Haroche, Phys. Rev. Lett. \textbf{79}, 769 (1997).

\bibitem{Duan2001}	 L. M. Duan, M. D. Lukin, J. I. Cirac, and P. Zoller, Nature (London) \textbf{414}, 413 (2001). 

\bibitem{Dayan2014} I. Shomroni, S. Rosenblum, Y. Lovsky, O. Bechler,
G. Guendelman, and B. Dayan, Science \textbf{345}, 903 (2014).
	 	
\bibitem{Domokos2002} P. Domokos, P. Horak, and H. Ritsch, Phys. Rev. A \textbf{65}, 033832 (2002).

\bibitem{Leuchs2007} M. Sondermann, R. Maiwald, H. Konermann, N. Lindlein, U.
Peschel, and G. Leuchs, Appl. Phys. B \textbf{89}, 489 (2007).

\bibitem{Leuchs2009} M. Stobi\'{n}ska, G. Alber, and G. Leuchs, Europhys. Lett. \textbf{86},
14007 (2009).

\bibitem{Fan2010} E. Rephaeli, J.-T. Shen, and S. Fan, Phys. Rev. A \textbf{82}, 033804
(2010).

\bibitem{Wang2011} Y. Wang, J. Min\'{a}\v{r}, L. Sheridan, and V. Scarani, Phys. Rev. A \textbf{83}, 063842 (2011).

\bibitem{Koenderink2011} Y. Chen, M.Wubs, J.M{\o}rk, and A. F. Koenderink, New J. Phys.
\textbf{13}, 103010 (2011).

\bibitem{Wang2012} Y. Wang, J. Min\'{a}\v{r}, and V. Scarani, Phys. Rev. A \textbf{86}, 023811 (2012).

\bibitem{GB} H. S. Rag and J. Gea-Banacloche, Phys. Rev. A \textbf{96}, 033817 (2017). 

\bibitem{Leuchs2010} S. Heugel, A.Villar,M. Sondermann, U. Peschel, and G. Leuchs,
Laser Phys. \textbf{20}, 100 (2010).

\bibitem{Leuchs2013} M. Bader, S. Heugel, A. L. Chekhov, M. Sondermann, and
G. Leuchs, New J. Phys. \textbf{15}, 123008 (2013).

\bibitem{Du2012} S. Zhang, C. Liu, S. Zhou, C.-S. Chuu, M. M. T. Loy, and
S. Du, Phys. Rev. Lett. \textbf{109}, 263601 (2012).

\bibitem{Kurtsiefer2013} S. A. Aljunid, G. Maslennikov, Y. Wang, H. L. Dao, V.
Scarani, and C. Kurtsiefer, Phys. Rev. Lett. \textbf{111}, 103001 (2013).

\bibitem{Martinis2014} J. Wenner, Y. Yin, Y. Chen, R. Barends, B. Chiaro, E. Jeffrey, J.
Kelly, A. Megrant, J. Y. Mutus, C. Neill, P. J. J. O’Malley,
P. Roushan, D. Sank, A. Vainsencher, T. C. White, A. N.
Korotkov, A. N. Cleland, and J. M. Martinis, Phys. Rev. Lett. \textbf{112}, 210501 (2014).

\bibitem{Du2014} C. Liu, Y. Sun, L. Zhao, S. Zhang, M. M. T. Loy, and S. Du,
Phys. Rev. Lett. \textbf{113}, 133601 (2014).

\bibitem{Kurtsiefer2016} V. Leong, M. A. Seidler, M. Steiner, A. Cer\`{e}, and C. Kurtsiefer,
Nat. Commun. \textbf{7}, 13716 (2016).

\bibitem{Jhe} H. Nha and W. Jhe, Phys. Rev. A \textbf{56}, 2213 (1997).

\bibitem{Klimov} V. V. Klimov and M. Ducloy, Phys. Rev. A \textbf{69}, 013812 (2004).

\bibitem{cesium decay} Fam Le Kien, S. Dutta Gupta, V. I. Balykin, and K. Hakuta, 
Phys. Rev. A \textbf{72}, 032509 (2005).
	
\bibitem{TongNat03} L. Tong, R. R. Gattass, J. B. Ashcom, S. He, J. Lou, M. Shen, I. Maxwell, and E. Mazur, Nature (London) \textbf{426}, 816 (2003).

\bibitem{review2016} For a review, see T. Nieddu, V. Gokhroo, and S. Nic Chormaic, J. Opt. \textbf{18}, 053001 (2016).

\bibitem{review2017} For another review, see P. Solano, J. A. Grover, J. E. Homan, S. Ravets, F. K. Fatemi, L. A. Orozco, and S. L. Rolston, Adv. At. Mol. Opt. Phys. \textbf{66}, 439 (2017).

\bibitem{Nayak2018} For a more recent review, see K. Nayak, M. Sadgrove, R. Yalla, F.~L.~Kien, and K. Hakuta, J. Opt. \textbf{20}, 073001 (2018).

\bibitem{Fam2014} Fam Le Kien and A. Rauschenbeutel, Phys. Rev. A \textbf{90}, 023805 (2014). 

\bibitem{Petersen2014} J. Petersen, J. Volz, and A. Rauschenbeutel, Science \textbf{346}, 67 (2014).

\bibitem{Mitsch14b} R. Mitsch, C. Sayrin, B. Albrecht, P. Schneeweiss, and A. Rauschenbeutel, Nature Commun. \textbf{5}, 5713 (2014).

\bibitem{sponhigh} Fam Le Kien, Th. Busch, Viet Giang Truong, and S. Nic Chormaic, Phys. Rev. A \textbf{96}, 043859 (2017).

\bibitem{Zeldovich} A. V. Dooghin, N. D. Kundikova, V. S. Liberman, and B. Y. Zeldovich, Phys. Rev. A \textbf{45}, 8204 (1992);
V. S. Liberman and B. Y. Zeldovich, Phys. Rev. A \textbf{46}, 5199 (1992); M. Y. Darsht, B. Y. Zeldovich, I. V. Kataevskaya, and N. D. Kundikova, JETP \textbf{80}, 817 (1995) [Zh. Eksp. Theor. Phys. \textbf{107}, 1464 (1995)].

\bibitem{Bliokh review} For a review, see K. Y. Bliokh, A. Aiello, and M. A. Alonso, in \textit{The Angular Momentum of Light}, edited by D. L. Andrews and M. Babiker
(Cambridge University Press, Cambridge, 2012), p. 174. 

\bibitem{Bliokh review2015} For a more recent review, see K. Y. Bliokh, F. J. Rodriguez-Fortu\~{n}o, F. Nori, and A. V. Zayats, Nature Photon. \textbf{9}, 796 (2015).

\bibitem{Bliokh2015} K. Y. Bliokh, D. Smirnova, and F. Nori, Science \textbf{348}, 1448 (2015).

\bibitem{Bliokh2014} K. Y. Bliokh, A. Y. Bekshaev, and F. Nori, Nature Commun. \textbf{5}, 3300 (2014).

\bibitem{Banzer review2015} For a review, see A. Aiello, P. Banzer, M. Neugebauer, and G. Leuchs, Nature Photon. \textbf{9}, 789 (2015).

\bibitem{Lodahl2017} P. Lodahl, S. Mahmoodian, S. Stobbe, P. Schneeweiss,
J. Volz, A. Rauschenbeutel, H. Pichler, and P. Zoller, Nature (London) \textbf{541}, 473 (2017).

\bibitem{fiber books} See, for example, 
D. Marcuse, \textit{Light Transmission Optics} 
(Krieger, Malabar, 1989);
A. W. Snyder and J. D. Love, \textit{Optical Waveguide Theory} (Chapman and Hall, New York, 1983).

\bibitem{Scully} M. O. Scully and M. S. Zubairy, \textit{Quantum Optics} (Cambridge
University Press, Cambridge, 1997).

\bibitem{Loudon} R. Loudon, \textit{The Quantum Theory of Light} (Oxford University Press, Oxford, 2000), p. 243.

\bibitem{highorder} Fam Le Kien, Th. Busch, Viet Giang Truong, and S. Nic Chormaic, Phys. Rev. A \textbf{96}, 023835 (2017).

\end{thebibliography}
\end{document}